\documentclass[a4paper,11pt]{article}

\usepackage{jheppub}

\usepackage{latexsym}
\usepackage{epsfig}
\usepackage{graphicx}
\usepackage[english]{babel}
\usepackage{amssymb}

\title{Exclusive $J/\psi$ production in ultraperipheral collisions at the LHC:  
constraints on the gluon distributions in the proton and nuclei}

\author{V. Guzey}
\author{and M. Zhalov}

\affiliation{National Research Center ``Kurchatov Institute'',
Petersburg Nuclear Physics Institute (PNPI), Gatchina, 188300, Russia}

\emailAdd{vguzey@pnpi.spb.ru}
\emailAdd{zhalov@pnpi.spb.ru}


\abstract{
Using the framework of collinear factorization in perturbative QCD (pQCD), 
we analyze the recent data of the 
LHCb and ALICE collaborations on exclusive photoproduction 
of $J/\psi$ in ultraperipheral $pp$ and $AA$ collisions,  
respectively. 
We demonstrate that the simultaneous analysis of the proton and Pb data
allows us to reduce the ambiguity of the pQCD description of the 
$\gamma p \to J/\psi p$ and $\gamma A \to J/\psi A$ cross sections and, hence, 
to place additional constraints on
the gluon distributions in the 
nucleon and nuclei at small $x$. 
We also make predictions for the cross section of
coherent exclusive photoproduction of $J/\psi$ in nucleus--nucleus 
ultraperipheral collisions 
accompanied by the electromagnetic excitation of nuclei and the subsequent neutron emission.
}

\begin{document} 
\maketitle
\flushbottom

\section{Introduction}

Recently the LHCb~\cite{lhcb} and ALICE~\cite{alice2,alice1} collaborations presented results 
of their measurements of 
exclusive $J/\psi$ photoproduction in ultraperipheral proton--proton and nucleus--nucleus
collisions  at the LHC, respectively. These results are of particular interest
because the analysis of exclusive $J/\psi$ photoproduction 
within the leading logarithmic approximation of perturbative QCD (pQCD) 
predicts~\cite{Ryskin:1992ui,Brodsky:1994kf,Ryskin:1995hz,Frankfurt:1995jw} 
that the cross section of this process is proportional to the gluon density of the target squared.
Since  the fraction $x$  of the target momentum carried by the gluons is inversely proportional to 
the beam momentum, at the high energies achievable at the LHC,
it is potentially possible to obtain unique information on the
small $x$ behavior of the gluon density in the proton and nuclei and, 
thus, to constrain the current ambiguity in this quantity.

The LHCb collaboration measured the yield of $J/\psi$ 
at forward rapidities ($2<y<4.5$) in proton--proton ultraperipheral collisions (UPCs) 
at 7 TeV~\cite{lhcb}. 
This data allowed one to extract the $\gamma p\to J/\psi p$ cross section 
in two regions of the $\gamma p$ center-of-mass energy $W_{\gamma p}$: (i) 
$0.4\, {\rm TeV} < W_{\gamma p} < 1.3\, {\rm TeV}$ corresponding to the values of 
$x=M^2_{J/\psi}/W^2_{\gamma p}$ in the range of
$6 \times 10^{-6}  < x < 6 \times 10^{-5}$, which significantly 
extends the region of small $x$ studied at HERA, and 
(ii) $15\, {\rm GeV} < W_{\gamma p} < 51$ GeV corresponding to $0.004 < x < 0.04$, 
which is covered by earlier high-precision fixed-target experiments.
The analysis~\cite{lhcb} confirmed the power law energy dependence
of the $\gamma p\to J/\psi p$ cross section ($\sigma (W_{\gamma p})\propto  W^{\delta}_{\gamma p}$ 
with 
$\delta=0.92 \pm 0.15$) 
consistent with the previous HERA results and
did not reveal any evidence of new phenomena such as the
onset of the gluon saturation regime at small $x$.

The ALICE collaboration measured coherent $J/\psi$ 
photoproduction in ultraperipheral PbPb collisions
at $\sqrt{s_{NN}} =2.76$ TeV. 
The cross section was measured in two regions of the rapidity of produced $J/\psi$: 
in the range of $-3.6<y<-2.6$~\cite{alice2} and at central rapidities of
$-0.9<y<0.9$~\cite{alice1}. 
The  $\gamma Pb \to J/\psi Pb$ cross section obtained in the data analysis
was compared~\cite{alice2,alice1} to different model predictions and
the better agreement was found with models  including moderate nuclear 
gluon shadowing (the predictions referred to as RSZ-LTA~\cite{rsz} and
AB-EPS09~\cite{ba2}, see figure~3 of~\cite{alice2} and figure~5 of~\cite{alice1}).

In this article, using the framework of collinear factorization of pQCD and the
 leading logarithmic approximation (LLA), 
we extend and continue the analysis of $J/\psi$ photoproduction in heavy ion UPCs~\cite{Guzey:2013xba}
and calculate
the $\gamma p \to J/\psi p$ and $\gamma Pb \to J/\psi Pb$ cross sections employing various
parameterizations of the gluon distribution in the proton and Pb and compare them to the data mentioned
above. We argue that the {\it simultaneous} description of the proton and Pb data is
(i) possible, (ii) points to sizable nuclear gluon shadowing
at $x = 10^{-3}$ and below, and (iii) places constraints on 
the value of the hard scale $\mu^2$  probed in coherent $J/\psi$ photoproduction, 
on the size of higher order corrections to the commonly used 
LLA formula suggested in~\cite{Ryskin:1992ui,Brodsky:1994kf,Ryskin:1995hz,Frankfurt:1995jw},
and, thus, on the gluon distribution in Pb down to
$x=10^{-3}$ 
as well as on the gluon distribution in the proton down to $x=6 \times 10^{-6}$.
These values provide an order of magnitude improvement in the small $x$ kinematic coverage
compared to those 
achieved respectively at HERA in lepton--proton  deep inelastic scattering (DIS) experiments and at 
fixed-target nuclear DIS experiments.

We also make predictions for the cross section of
coherent exclusive photoproduction of $J/\psi$ in nucleus--nucleus 
UPCs accompanied by the electromagnetic excitation of nuclei and the subsequent neutron emission.

\section{UPCs at the LHC and exclusive $J/\psi$ photoproduction in pQCD}
\label{sec:AA}

\subsection{Exclusive $J/\psi$ photoproduction on the proton and nuclei in LLA of pQCD}
\label{subsec:pQCD_proton}

The cross section of $J/\psi$ production in symmetric nucleus--nucleus UPCs 
(for a review of the UPC physics, see~\cite{Baltz:2007kq}) has the following form: 
\begin{eqnarray}
 \frac{d \sigma_{AA\to AAJ/\psi}(y)}{dy} 
=N_{\gamma/A}(y)\sigma_{\gamma A\to J/\psi A}(y)+
N_{\gamma/A}(-y)\sigma_{\gamma A\to J/\psi A}(-y) \,,
\label{csupc}
\end{eqnarray}
where $N_{\gamma/A}(y)=\omega dN_{\gamma/A}(\omega)/d\omega$ is the photon flux; 
$y = \ln(2\omega/M_{J/\psi})$ is the $J/\psi$ rapidity, where $\omega$ is the photon 
energy and $M_{J/\psi}$ is the mass of $J/\psi$.
 The presence of two terms in eq.~(\ref{csupc}) reflects the specific feature of UPCs: 
each colliding nucleus can radiate a photon and can also serve as a target. 
Thus, at each value of the rapidity $y$---except for the case of $y=0$---there are two 
contributions corresponding to production of $J/\psi$ by low energy photons with the energy of 
$\omega_{L}=(M_{J/\psi}/2)e^{-y}$ and by high energy photons with the energy of
$\omega_{H}=(M_{J/\psi}/2)e^{y}$, respectively. 
The separation of these two contributions can be easy realized
in two cases: (i)
at $y=0$, when one obtains
$\sigma_{\gamma A\to J/\psi A}(y=0)=(d\sigma_{AA\to AAJ/\psi}(y=0)/dy)/(2N_{\gamma/A}(y=0))$,
and (ii) in the region of the rapidities $y$, where one of the contributions dominates and, hence, 
one obtains $\sigma_{\gamma A\to J/\psi A}(y)=(d\sigma_{AA\to AAJ/\psi}(y)/dy)/N_{\gamma/A}(y)$. 
Therefore, since the photon flux $N_{\gamma/A}(y)$ can be calculated with 
reasonable accuracy (with the uncertainty of the order of $5\%$), $J/\psi$ production in UPCs 
can be effectively used to study the 
$J/\psi$
photoproduction cross section.

At high energies and small transverse momenta $p_{t}$ of $J/\psi$ ($W_{\gamma p} \gg 
M_{J/\psi} \gg p_t$), using the framework of collinear factorization and in
the leading logarithmic approximation (LLA) of pQCD which is usually referred to
as the leading order (LO) approximation,
the cross section of coherent $J/\psi$ photoproduction on 
the proton or a nucleus ($T=p,A$) 
reads~\cite{Ryskin:1992ui,Brodsky:1994kf,Ryskin:1995hz,Frankfurt:1995jw,Frankfurt:1997fj}:
\begin{eqnarray}
\frac {d\sigma_{\gamma T\rightarrow J/\psi T}(W_{\gamma p},t=0)}{dt}=
\frac{M^3_{J/\psi}\Gamma_{ee}{\pi}^3}{48\,\alpha_{\rm e.m.} \mu^8} (1+\eta^2)R_g^2 \,F^2(\mu^2)
\left[\alpha_s(\mu^2) x G_T (x,\mu^2 )\right]^2 \,,
\label{dcslo}
\end{eqnarray} 
where
$\Gamma_{ee}$ is the width of the $J/\psi$
electronic decay; $\alpha_{\rm em}$ is the fine structure constant; $\alpha_s(\mu^2)$ is the 
running strong coupling constant.
In eq.~(\ref{dcslo}),
$G_T(x,\mu^2)$ is the density of gluons 
in the target $T$, where the gluons carry the momentum fraction $x=M_{J/\psi}^2/W_{\gamma p}^2$ 
of the target and are resolved at the hard scale $\mu$ which is of the order of the mass 
of the charm quark, $m_c$ (the precise value of 
$\mu^2$ will be discussed in section~\ref{subsec:proton}).

Equation~(\ref{dcslo}) contains several factors taking into account intricate aspects of
the discussed reaction which mostly affect the normalization of the $\gamma p \to J/\psi p$ cross section,  
but not so much its $W_{\gamma p}$ dependence. 
\begin{itemize}

\item
The factor of $\eta$ is the ratio of the real to the imaginary parts of the 
$\gamma T \to J/\psi T$ amplitude.
This quantity can be evaluated using  the well-known 
Gribov--Migdal relation~\cite{Gribov:1968uy,Martin:2007sb}, 
$\eta =\tan\left(\pi \lambda/2\right)$,
where $\lambda$ parametrizes the behavior of $xG_T (x,\mu^2)$ at small $x$, 
$xG_T (x,\mu^2) \propto 1/x^{\lambda}$. 
Taking $\lambda=0.2$, 
which broadly reproduces the energy behavior of the $\gamma p \to J/\psi p$  cross section 
(see the discussion below), we obtain $1+\eta^2=1.11$.

\item
In the framework of collinear QCD factorization for hard exclusive 
processes~\cite{Collins:1996fb,Collins:1998be}, the $\gamma p \to J/\psi p$ amplitude is
expressed in terms of the gluon {\it generalized} parton distribution (GPD) rather than the usual one. 
One can take this fact into account phenomenologically by introducing the enhancement factor 
$R_g$~\cite{Martin:2007sb,Shuvaev:1999ce,Harland-Lang:2013xba}:
\begin{equation}
R_g=\frac{2^{3+2 \lambda}}{\sqrt{\pi}} \frac{\Gamma(\frac{5}{2}+\lambda)}{\Gamma(4+\lambda)} \,.
\label{eq:R_g}
\end{equation}
Note that the expression for $R_g$ depends on the model/parameterization for the gluon GPDs and, hence, 
is model-dependent.
To estimate the magnitude of the effect, we use $\lambda=0.2$ and obtain 
$R_g=1.18$~\cite{Martin:2007sb}.

\item 
Note that in our analysis, we employed the form $xG_T (x,\mu^2) \propto 1/x^{\lambda}$ only
to evaluate the factors of $\eta$ and $R_g$. Moreover, in the above estimates, we used 
$\lambda=0.2$ only for illustration. 
In practice, in the interval $5 \times 10^{-6} \leq x \leq 0.01$, each gluon 
distribution used in our work (including nuclear gluon distributions)
was fitted using the shape $xG_T (x,\mu^2) \propto 1/x^{\lambda}$ and
the corresponding $\lambda$ was determined from the fit.
We found that this simple fit works with a high accuracy and the resulting values of $\lambda$ can
be taken constant (except for the MSTW08 gluon distribution) and are 
close to $\lambda=0.2$ (for details, see the Appendix and table~\ref{table:lambda} there). 

In the $0.01 \leq x \leq 0.1$ interval, which corresponds to the low-$W$ and large-$|y|$ edges of 
the kinematics considered in our work,
 the gluon distributions can no longer be accurately
approximated by the simple fit $xG_T (x,\mu^2) \propto 1/x^{\lambda}$ 
(except for the MNRT07 gluon distribution). Therefore, 
the small-$x$ asymptotic expressions of eq.~(\ref{eq:R_g}) and $\eta =\tan\left(\pi \lambda/2\right)$ 
become progressively less accurate.
Our analysis of $R_g$ based on~\cite{Harland-Lang:2013xba} and of $\eta$ based on~\cite{Bronzan:1974jh}
shows that the extrapolation of our low-$x$ results to the  $0.01 \leq x \leq 0.1$ interval
leads to certain theoretical uncertainties which increase with an increase of $x$.
Depending on the used gluon distribution, these uncertainties 
are, e.g., of the order of $5-10$\% for $R_g$ and $5-7$\% for $1+\eta^2$ at $x=0.05$
and of the order of
$10-20$\% for $R_g$ and $10-15$\% for $1+\eta^2$ at $x=0.1$.

\item
The main uncertainty in eq.~(\ref{dcslo}) comes from
the suppression factor $F^2(\mu^2)$,
 $0 <F^2(\mu^2) < 1$, which contains all effects that go beyond the leading order collinear factorization
that we use: next-to-leading order corrections, the overlap between the photon and 
$J/\psi$ light-cone wave functions weighted by the gluon distribution of the target either in the $k_{T}$ 
factorization approach~\cite{Ryskin:1995hz} or in the dipole 
formalism~\cite{Frankfurt:1995jw,Frankfurt:1997fj,Frankfurt:2000ez}, 
corrections associated with the charmonium wave function~\cite{hoodbhoy}, etc.

\end{itemize}

Equation~(\ref{dcslo}) was originally derived in the non-relativistic approximation for the 
charmonium wave function~\cite{Ryskin:1992ui} that neglects the motion of the charm
quarks in $J/\psi$. In this case, $F^2(\mu^2)=1$ and $\mu^2=M_{J/\psi}^2/4=2.4$ GeV$^2$.
The magnitude of relativistic and other corrections, i.e., the deviation of $F^2(\mu^2)$ from unity, and
the related issue of the choice of the value of $\mu^2$ are subjects of 
discussion in the literature~\cite{Ryskin:1995hz,Frankfurt:1997fj,Frankfurt:2000ez,hoodbhoy}.
However, while these corrections mainly influence the absolute value of the $\gamma p\to J/\psi p$ 
cross section 
(the total uncertainty of LO pQCD predictions is estimated~\cite{Ryskin:1995hz,hoodbhoy} to be about
$30\%$ or less), its
energy dependence is predicted by eq.~(\ref{dcslo}) and is given by the $x$ dependence
of $xG_T(x,\mu^2)$.

In this work, we do not attempt to evaluate $F^2(\mu^2)$ theoretically using, e.g., different models
of the charmonium wave function. Instead, we adopt the following strategy.
First, by comparing predictions of eq.~(\ref{dcslo}) with the available high-energy data
on the $W_{\gamma p}$ dependence of the $\gamma p\to J/\psi p$ cross section, 
we fix the value of $\mu^2$. Second, by requiring that eq.~(\ref{dcslo}) describes also 
the magnitude of the $\gamma p\to J/\psi p$ cross section, we determine the normalization
of the LO pQCD cross section given by eq.~(\ref{dcslo}), i.e., we phenomenologically find
 $F^2(\mu^2)$.

Turning to the $t$ dependence of the $d \sigma_{\gamma p\to J/\psi p}/dt$ cross section, 
we note that its
$p_{t}$ dependence is
usually parameterized in the exponential form of
$\exp [-B_{J/\psi}(W_{\gamma p})p^{2}_{t}]$, 
with the slope parameter $B_{J/\psi}(W_{\gamma p})$ weakly depending on energy.
A good description of the HERA data on the $t$ dependence of the $J/\psi$ 
photoproduction cross section~\cite{H1_2000,ZEUS_2002,H1_2005,H1_2013} 
is obtained using the following Regge-motivated 
parameterization of the slope $B_{J/\psi}(W_{\gamma p})$:
\begin{eqnarray} 
B_{J/\psi}(W_{\gamma p})=
4.5+0.4\ln\left(\frac {W_{\gamma p}}{90 \,{\rm GeV}}\right) \,,
\label{slope}
\end{eqnarray} 
which is consistent with the parameterizations and the values of 
$B_{J/\psi}(W_{\gamma p})$
reported in~\cite{H1_2000,ZEUS_2002,H1_2005,H1_2013}.

Hence, for the proton target, the total $J/\psi$ photoproduction cross section is: 
\begin{eqnarray}
\sigma_{\gamma p\to J/\psi p}(W_{\gamma p})=
\frac{M^3_{J/\psi}\Gamma_{ee}{\pi}^3}{48\,\alpha_{\rm e.m.}\mu^8} \frac{1}{B_{J/\psi}(W_{\gamma p})}
(1+\eta^2)\,F^2(\mu^2)
\left[R_g \alpha_s(\mu^2) xG_p(x,\mu^2 )\right]^2 \,,
\label{csprot}
\end{eqnarray} 
where we ignored the negligibly small effect of the minimal momentum transfer to the target.  

Extending eqs.~(\ref{dcslo}) and (\ref{csprot}) to the description of $J/\psi$ production on nuclei and 
accounting for the transverse momentum distribution dictated by the nuclear form 
factor $F_A (t)$, one obtains~\cite{fgsz}: 
\begin{eqnarray}
\sigma_{\gamma A \rightarrow J/\psi A}(W_{\gamma p})&=&
\frac{M^3_{J/\psi}\Gamma_{ee}{\pi}^3}{48\,\alpha_{\rm e.m.}\mu^8} (1+\eta_A^2)\,F^2(\mu^2)
\left[\frac {1}{A}R_{g,A}\, \alpha_s(\mu^2) xG_A(x,\mu^2)\right]^2 \Phi_A(t_{\rm min})= 
\nonumber \\
&=& \frac{(1+\eta_A^2) R_{g,A}^2}{(1+\eta^2) R_{g}^2}\,\frac{d\sigma_{\gamma p\rightarrow J/\psi p}(W_{\gamma p},t=0)}{dt}
\left[\frac {G_A(x,{\mu}^2)} {AG_N(x,{\mu}^2)} \right]^2 {\Phi_A(t_{\rm min})} \,,
\label{csa1} 
\end{eqnarray}
where $G_A(x,{\mu}^2)$ is the nuclear gluon density;  
$\Phi_A(t_{\rm min})=\int \limits_{t_{\rm min}}^{\infty} dt \left|F_A (t)\right|^2$,
where $t_{\rm min}=-M_{J/\psi}^4 m_N^2/W_{\gamma p}^4$ is the minimal momentum transfer to the nucleus;
$\eta_A$ is the ratio of the real to the imaginary parts of 
the $\gamma A \to J/\psi A$ scattering amplitude whose skewness effect is parameterized by
the factor of $R_{g,A}$. 
By analogy with the free proton case
(see the discussion above and eq.~(\ref{eq:R_g})),
we determine the factors of $\eta_A$ and $R_{g,A}$ using the asymptotic small-$x$ dependence of
the nuclear gluon distribution $xG_A(x,\mu^2) \propto 1/x^{\lambda_A}$: $\eta_A =\tan\left(\pi \lambda_A/2\right)$ and 
$R_{g,A}=(2^{3+2 \lambda_A}/\sqrt{\pi}) \Gamma(\frac{5}{2}+\lambda_A)/\Gamma(4+\lambda_A)$.

In this work, we are interested in small values of $x$ probed in high-energy 
$J/\psi$ photoproduction experiments, 
 $6 \times 10^{-5} < x < 0.05$,
where the effect of nuclear shadowing 
significantly reduces the nuclear gluon density $G_A(x,\mu^2)$ compared to the free nucleon
gluon density $G_N(x,\mu^2)$, i.e., $G_A(x,\mu^2) < A G_N(x,\mu^2)$.
Hence, within LO pQCD, the influence of the nuclear medium on coherent $J/\psi$ 
photoproduction on nuclei [see eq.~(\ref{csa1})] is related to the nuclear gluon shadowing characterized 
by the factor
\begin{eqnarray}
R(x,\mu^2)=\frac{G_A(x,{\mu}^2)}{AG_N(x,{\mu}^2)} \,.
\label{shad}
\end{eqnarray}

In addition, nuclear shadowing tames the growth of the nuclear gluon density $G_A(x,\mu^2)$ 
 with a decrease of $x$ and makes it slower than that of the free nucleon. 
Parameterizing $G_A(x,\mu^2)$ at small $x$ as $G_A(x,\mu^2) \propto 1/x^{\lambda_A}$, 
one finds that $\lambda_A < \lambda_p$ ($\lambda_p$ corresponds to the free proton) and, hence, 
\begin{equation}
\kappa_{A/N} \equiv \left[\frac{(1+\eta_A^2) R_{g,A}^2}{(1+\eta^2) R_{g}^2} \right]^{1/2}= 0.87-0.97 < 1 \,,
\label{eq:dop_supp}
\end{equation}
where the spread in the values of $\kappa_{A/N}$ originates from different $x$ dependences of 
various nuclear and free proton gluon distributions that we used in this work.
Thus, in addition to the factor of $R(x,\mu^2)$, the factor of $\kappa_{A/N}$ 
further decreases the $\gamma A \to J/ \psi A$ cross section 
compared to the $\gamma p \to J/ \psi p$ one.

It is important to point out---and it is one of the key points of the present 
work---that consistency of the simultaneous LO pQCD description
of the $\gamma A \to J/ \psi A$ and $\gamma p \to J/ \psi p$ cross sections, i.e., 
consistency of eqs.~(\ref{csprot}) and (\ref{csa1}), 
requires the use of the same proton gluon density $G_p(x,\mu^2)$ in  
the forward $\gamma p\to J/\psi p$  cross section and in the evaluation of 
the nuclear gluon shadowing factor $R(x,\mu^2)$.

\subsection{LO pQCD description of $J/\psi$ photoproduction on the proton}
\label{subsec:proton}

The results of the calculation of the total $\gamma p\to J/\psi p$ cross section 
using eq.~(\ref{csprot}) and different gluon parton distributions in the proton 
(see the Appendix) 
are presented in figure~\ref{fig:Sigma_proton_pQCD_Q2_2_4}. 
These results are compared to
the available high-energy HERA data on the $\gamma p \to J/\psi p$ cross section (we only omitted
the 2000 H1 data~\cite{H1_2000} which is consistent with the more recent data), the  
$\gamma p \to J/\psi p$ cross section extracted from the LHCb data on proton--proton UPCs~\cite{lhcb},
and the fit~\cite{Strikman:2005ze,Guzey:2013xba} to the available data on $d \sigma_{\gamma p \to J/\psi p}(W_{\gamma p},t=0)/dt$ converted into
the total cross section using the slope $B_{J/\psi}(W_{\gamma p})$ of eq.~(\ref{slope}).

In the upper panel of figure~\ref{fig:Sigma_proton_pQCD_Q2_2_4}, the theoretical predictions of 
eq.~(\ref{csprot}) are evaluated with $F^2(\mu^2)=1$ and at $\mu^2=M_{J/\psi}^2/4=2.4$ GeV$^2$~\cite{Ryskin:1992ui} 
 using the 
MNRT07~\cite{Martin:2007sb}, CTEQ6L1 and CTEQ6L~\cite{Pumplin:2002vw}, MRST04~\cite{Martin:2006qz}, 
NNPDF~\cite{Ball:2010de}, and MSTW08LO~\cite{Martin:2009iq}
leading order (LO) gluon parton distributions in the proton (the curves in the figure are 
labeled accordingly).
In the Appendix, we show these gluon distributions as a function of $x$ 
(figure~\ref{fig:Glue_proton_pQCD_Q2_3}) and also give the corresponding values of the strong 
coupling constant $\alpha_s$ entering eq.~(\ref{csprot}) (table~\ref{table:alpha_s}).
 One can see from 
the panel that while the MNRT07 parameterization provides an excellent description of the 
data---including the region of high $W_{\gamma p}$ measured by the LHCb collaboration which was not
used in the analysis of ref.~\cite{Martin:2007sb}---all other gluon distributions fail to reproduce 
either the normalization of the data or both the normalization and the energy dependence 
of the data. 

\begin{figure}[t]
\centering
\epsfig{file=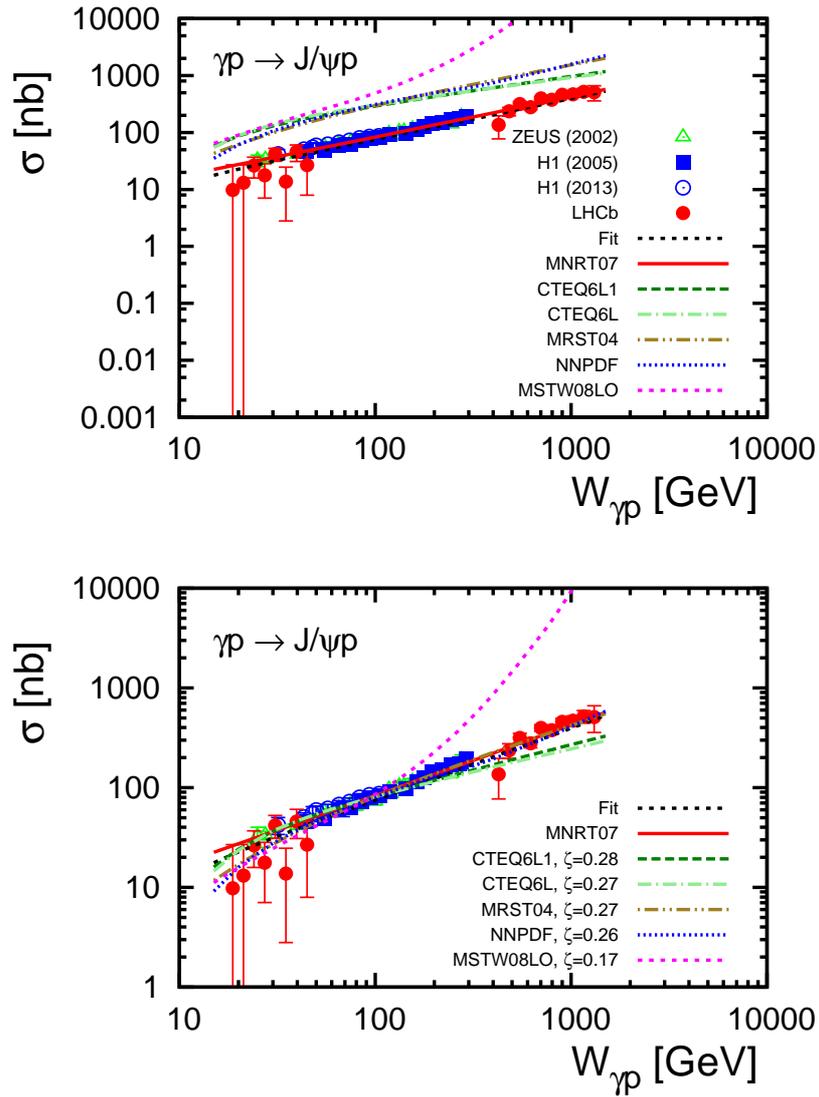,scale=1.7}
\caption{The $\gamma p\to J/\psi p$ cross section as a function of the $\gamma p$ center-of-mass energy
$W_{\gamma p}$. {\it Upper panel}: The available high-energy data vs.~the fit~\cite{Guzey:2013xba} 
(the curve labeled ``Fit'') and LO pQCD predictions of eq.~(\ref{csprot}) at $\mu^2=2.4$ GeV$^2$
using various gluon distributions of the proton, see text and the Appendix.
{\it Lower panel}: The same as in the upper panel, but with the pQCD predictions reduced 
by the multiplicative factor $\zeta=F^2(\mu^2)$~(\ref{eq:zeta}).
}
\label{fig:Sigma_proton_pQCD_Q2_2_4}
\end{figure}

To illustrate this point, we introduce the normalization factor $\zeta$,
\begin{equation}
\frac{1}{\zeta}=\frac{\sigma_{\gamma p\to J/\psi p}(W_{\gamma p}=100 \, {\rm GeV})_{|{\rm Eq}.~(\ref{csprot})}}{81 \, {\rm mb}} \,,
\label{eq:zeta}
\end{equation} 
which is designed to reduce the value of the theoretically calculated cross section at 
$W_{\gamma p}=100$ GeV so that it reproduces the data at this point.
One can see that $\zeta$ plays the role of the suppression factor $F^2(\mu^2)$ that we discussed above.
The second column of table~\ref{table:zeta} gives the resulting values of $\zeta$ at $\mu^2=2.4$ GeV$^2$.
\begin{table}[htbp]
  \centering
  \begin{tabular}{|c|c|c|}
    \hline
Parameterization & $\zeta(Q^2=2.4 \ {\rm GeV}^2)$ &  $\zeta(Q^2=3 \ {\rm GeV}^2)$\\
\hline
CTEQ6L1~\cite{Pumplin:2002vw} & $0.28$   & $0.48$  \\
CTEQ6L~\cite{Pumplin:2002vw}  & $0.27$   & $0.49$ \\
MRST04~\cite{Martin:2006qz}   & $0.27$   & $0.50$  \\
NNPDF~\cite{Ball:2010de}      & $0.26$   & $0.52$  \\
MSTW08LO~\cite{Martin:2009iq} & $0.17$   & $0.30$  \\
\hline
\end{tabular}
\caption{The normalization factor $\zeta$ of eq.~(\ref{eq:zeta}).}
\label{table:zeta}
\end{table}

The lower panel of figure~\ref{fig:Sigma_proton_pQCD_Q2_2_4} shows the LO pQCD predictions of 
eq.~(\ref{csprot}) at $\mu^2=2.4$ GeV$^2$ multiplied by the corresponding factor of $\zeta$.
One can see from the figure that after the renormalization by the $\zeta$ factor, 
the predictions using the MRST04 and NNPDF gluon distributions correctly reproduce the energy 
behavior and the normalization of the $\gamma p\to J/\psi p$ cross section over the considered $W_{\gamma p}$
range. At the same time, the predictions with the CTEQ6L1, CTEQ6L and MSTW08LO parameterizations
fail to reproduce the energy behavior of
the $\gamma p\to J/\psi p$ cross section for $W_{\gamma p} > 100$ GeV: the behavior predicted by the 
MSTW08LO fit is distinctly too steep, i.e., the resulting values of  $\sigma_{\gamma p\to J/\psi p}$
contradict the LHCb data,
 and the behavior predicted by the 
CTEQ6L1 and CTEQ6L fits is too slow.

In addition, as we will demonstrate in section~\ref{subsec:nucleus}, the MNRT07 gluon distribution
at $\mu^2=2.4$ GeV$^2$ corresponds to the prediction made in the framework of the leading twist 
nuclear shadowing, which overestimates the nuclear suppression factor $S(W_{\gamma p})$ at $x=10^{-3}$
(see figure~\ref{fig:S_pb208_Q2_2_4}). 

We have already mentioned above the uncertainty of the LO pQCD calculation of 
$\sigma_{\gamma p\to J/\psi p}$ related to the uncertainty in the choice of $\mu^2$ 
in eq.~(\ref{csprot}). In particular, 
on the one hand, the non-relativistic approximation that neglects the transverse momenta (Fermi motion)
of $c$-quarks  in the $J/\psi$ wave function prescribes that $\mu^2=M_{J/\psi}^2/4=2.4$ 
GeV$^2$~\cite{Ryskin:1992ui}.
On the other hand, the scale $\mu^2$ cannot be 
reliably
 fixed by the LO formalism.
The analyses accounting for effects that are not included in our formalism (see the discussion of the
factor of $F^2(\mu^2)$ above) suggest that the appropriate hard scale of the $\gamma p\to J/\psi p$ process
is $\mu^2 > 2.4$ GeV$^2$. 

We approach this problem phenomenologically 
by increasing $\mu^2$ in eq.~(\ref{csprot}) 
from $\mu=2.4$ GeV$^2$ to a certain value of $\mu^2$ so that  eq.~(\ref{csprot}) correctly 
reproduces the energy behavior of the data in figure~\ref{fig:Sigma_proton_pQCD_Q2_2_4}, which corresponds to
$x G_p(x,\mu^2) \propto 1/x^{\lambda}$ with $\lambda \approx 0.2$.
Note that since the cross section is proportional to $1/\mu^8$, even a rather small increase of $\mu^2$ 
leads to a sizable reduction of the $\gamma p\to J/\psi p$ cross section.

For small $x$, the MSTW08LO fit corresponds to $\lambda \gg 0.2$ already at $\mu^2=2.4$ GeV$^2$. 
Therefore, 
an increase of $\mu^2$ will further increase $\lambda$ and, thus, will make the description of the data by the 
corresponding LO pQCD calculation only worse (see the lower panel of 
figure~\ref{fig:Sigma_proton_pQCD_Q2_2_4}).
This forces us to conclude that the rapid increase of the MSTW08LO gluon distribution for small $x$
contradicts the consistent simultaneous description of the HERA and LHCb data on 
exclusive $J/\psi$ photoproduction on the proton
(see also figure~\ref{fig:Glue_proton_pQCD_Q2_3_comparison} in the Appendix).

\begin{figure}[th]
\centering
\epsfig{file=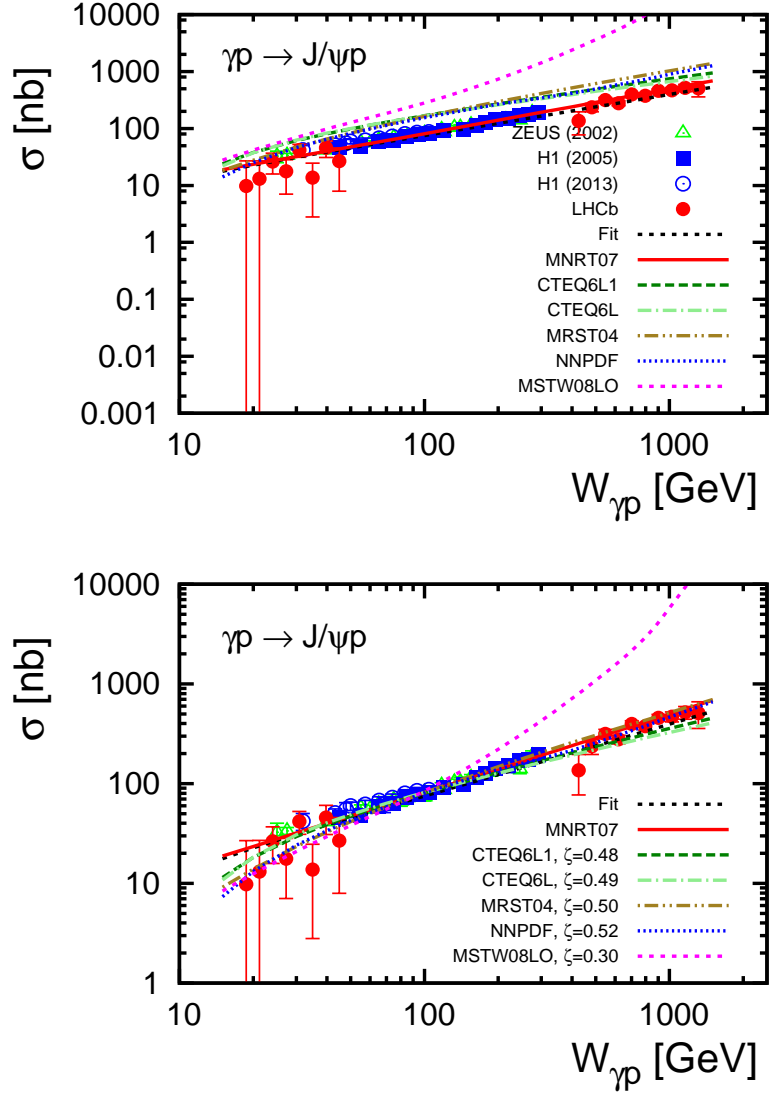,scale=1.7}
\caption{The same as in figure~\ref{fig:Sigma_proton_pQCD_Q2_2_4}, but with the LO pQCD predictions
evaluated at $\mu^2=3$ GeV$^2$.}
\label{fig:Sigma_proton_pQCD_Q2_3_0}
\end{figure}

Our analysis shows that we can obtain a good description of the energy dependence of the data presented
in figure~\ref{fig:Sigma_proton_pQCD_Q2_2_4} by eq.~(\ref{csprot}) evaluated at $\mu^2=3$ GeV$^2$ and 
this description will be equally good for all tested gluon distributions in the proton, except for 
 MSTW08LO which has already been ruled out. This is illustrated in 
figure~\ref{fig:Sigma_proton_pQCD_Q2_3_0}, where the upper panel presents predictions of eq.~(\ref{csprot})
at  $\mu^2=3$ (assuming $F^2(\mu^2)=1$) and the lower panel presents the same predictions multiplied
by the corresponding factor of $\zeta$, which plays the role of $F^2(\mu^2)$.
The corresponding values of  $\zeta$ at $\mu^2=3$ GeV$^2$ are given in the third column of 
table~\ref{table:zeta}.
One can see from the lower panel of figure~\ref{fig:Sigma_proton_pQCD_Q2_3_0} that the choice of
$\mu^2=3$ GeV$^2$ and $F^2(\mu^2) \approx 0.5$ in eq.~(\ref{csprot}) allows one to achieve the good
description of the $W_{\gamma p}$ dependence and normalization of the $\gamma p\to J/\psi p$ cross section
at collider energies using various sets of the gluon distribution in the proton.

Note that the MNRT07 prediction at $\mu^2=3$ GeV$^2$ in figure~\ref{fig:Sigma_proton_pQCD_Q2_3_0} 
has been scaled by the appropriate factor to reproduce the normalization of the data---we chose not
to show explicitly this normalization (the corresponding $\zeta=1.9$) because the normalization 
of this gluon distribution is obtained from a fit to the data.

\subsection{LO pQCD description of $J/\psi$ photoproduction on Lead}
\label{subsec:nucleus}  

It was shown in~\cite{Guzey:2013xba} that the recent ALICE measurements of exclusive $J/\psi$ production
in ultraperipheral PbPb collisions~\cite{alice2,alice1} provide direct evidence for strong nuclear 
gluon shadowing down to $x \sim 10^{-3}$, which can be model-independently quantified by
the factor of $S(W_{\gamma p})$:
\begin{equation}
S(W_{\gamma p})=\left[\frac{\sigma^{\rm exp}_{\gamma Pb \to J/\psi Pb}(W_{\gamma p})}{\sigma^{\rm IA}_{\gamma Pb \to J/\psi Pb}((W_{\gamma p})} \right]^{1/2} \,.
\label{eq:S}
\end{equation}
In eq.~(\ref{eq:S}), the numerator is the experimental cross section and the denominator is the result of the 
impulse approximation neglecting all nuclear effects except for coherence, i.e.,
the fact that $\Phi_A(t_{\rm min}) \neq 1$.
 The analysis of~\cite{Guzey:2013xba} 
found that $S(x=0.025)=0.74 \pm 0.12$ and $S(x=0.001)=0.61 \pm 0.05$.
(We used $x=M_{J/\psi}^2/W_{\gamma p}^2$.)

Combining Eqs.~(\ref{dcslo}) and (\ref{csa1}), we obtain
\begin{equation}
S(W_{\gamma p})=\kappa_{A/N}\, \frac{G_A(x,{\mu}^2)}{AG_N(x,{\mu}^2)}=\kappa_{A/N}\,R(x,\mu^2) \,,
\label{eq:S_2}
\end{equation}
where $R(x,\mu^2)=G_A(x,{\mu}^2)/[AG_N(x,{\mu}^2)]$ is the factor characterizing nuclear shadowing in the gluon
channel and $\kappa_{N/A}$ is given by eq.~(\ref{eq:dop_supp}).

\begin{figure}[h]
\centering
\epsfig{file=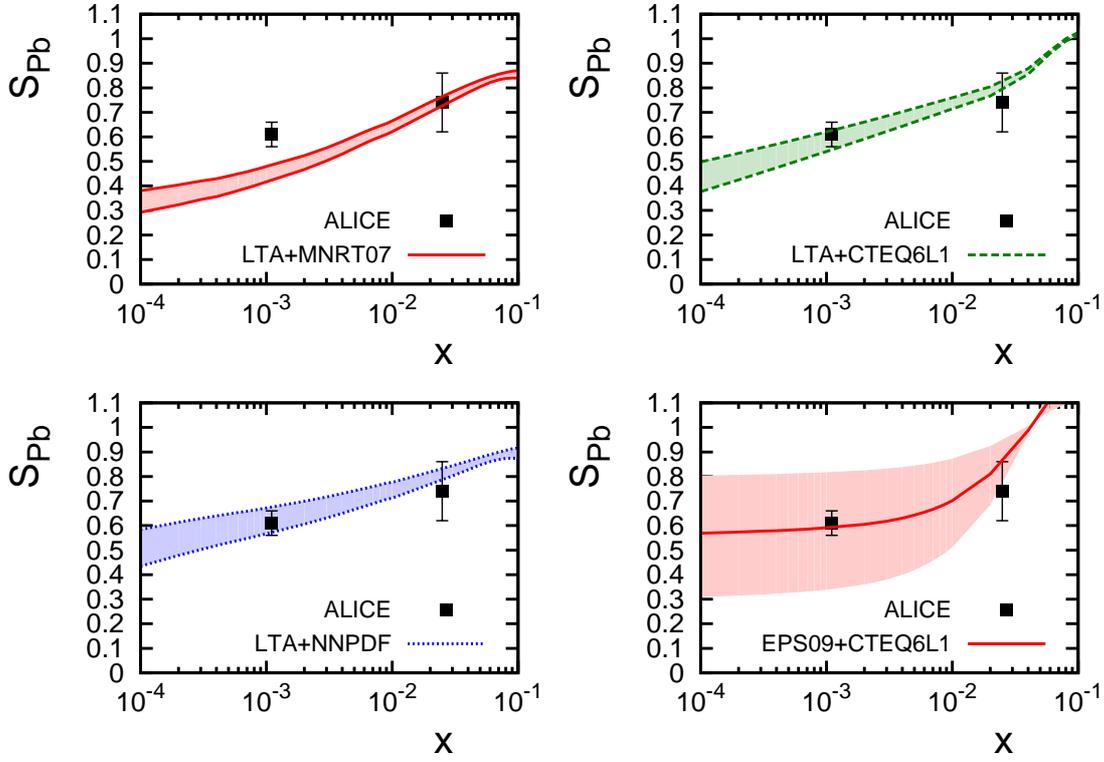,scale=1.45}
\caption{The suppression factor $S(W_{\gamma p})$ for Lead of eq.~(\ref{eq:S})
as a function of $x=M_{J/\psi}^2/W_{\gamma p}^2$. 
The results of the model-independent analysis of the ALICE data of~\cite{Guzey:2013xba} 
(labeled ``ALICE'') are
compared with the LO pQCD predictions at $\mu^2=2.4$ GeV$^2$ (see text for details).
}
\label{fig:S_pb208_Q2_2_4}
\end{figure}

\begin{figure}[h]
\centering
\epsfig{file=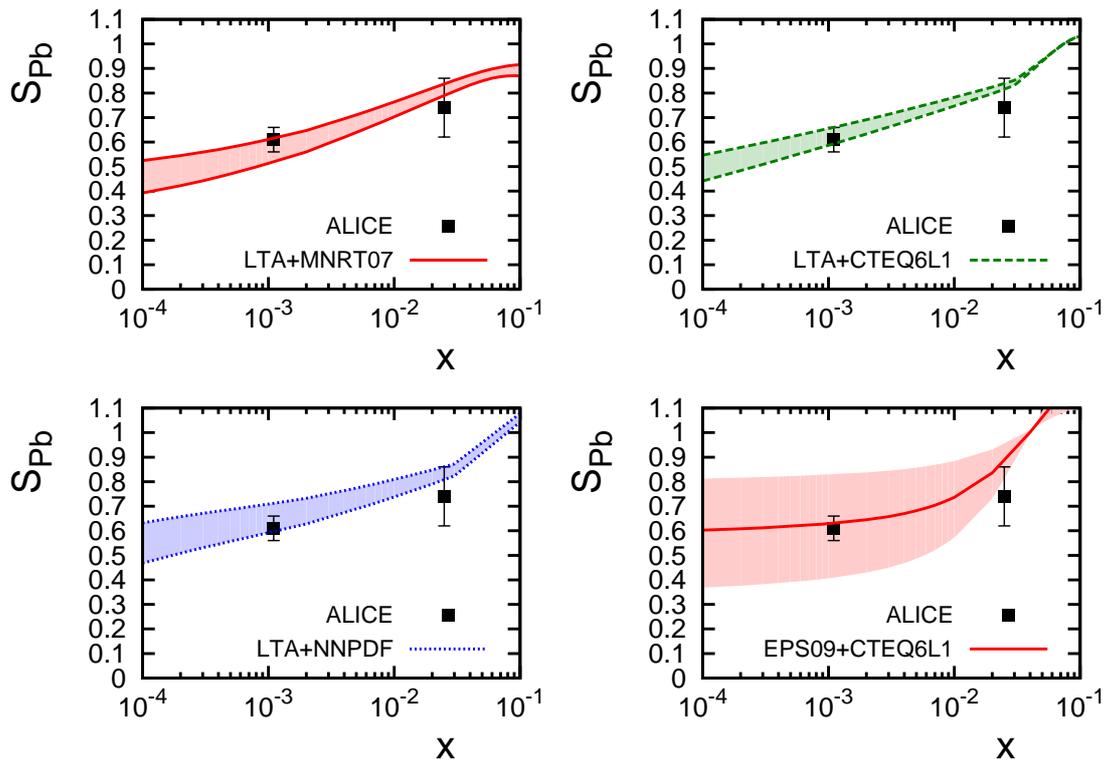,scale=1.45}
\caption{The same as in figure~\ref{fig:S_pb208_Q2_2_4}, but with the LO pQCD predictions evaluated
at $\mu^2=3$ GeV$^2$.
}
\label{fig:S_pb208_Q2_3_0}
\end{figure}

Figures~\ref{fig:S_pb208_Q2_2_4} and \ref{fig:S_pb208_Q2_3_0} present 
the suppression factor $S(W_{\gamma p})$ for Lead  as a function of $x=M_{J/\psi}^2/W_{\gamma p}^2$.
The results of the model-independent analysis of the ALICE data of~\cite{Guzey:2013xba}
(labeled ``ALICE'')
are compared with the 
LO pQCD predictions given by eq.~(\ref{eq:S_2}) at $\mu^2=2.4$ GeV$^2$ 
(figure~\ref{fig:S_pb208_Q2_2_4}) and at $\mu^2=3$ GeV$^2$ (figure~\ref{fig:S_pb208_Q2_3_0}).
In the two upper panels and in the lower left one,
the factors of $R(x,\mu^2)$ and $\kappa_{A/N}$ are calculated in the framework of the 
leading twist approximation (LTA) consisting in the combination of the leading twist theory of 
nuclear shadowing~\cite{Frankfurt:2011cs} with the given (MNRT07, CTEQ6L1, CTEQ6L, MRST04 and NNPDF) 
gluon distributions of the free nucleon.
In each case, we show the band of predictions which corresponds to the intrinsic uncertainty of the leading
twist theory of nuclear shadowing\footnote{The bands shown in figures~\ref{fig:S_pb208_Q2_2_4} and
\ref{fig:S_pb208_Q2_3_0} represent the
theoretical uncertainty of the leading twist theory of nuclear shadowing~\cite{Frankfurt:2011cs}
associated with the ambiguity in the magnitude of the contribution describing the
interaction of the virtual photon with three and more nucleons of the nucleus. 
The upper and lower boundaries of the bands correspond to the lower and higher limits on shadowing.}.
Note also that since the predictions with the CTEQ6L1 and CTEQ6L and with the MRST04 and NNPDF gluon
distributions are rather close, we show only the representative examples of CTEQ6L1 and NNPDF.

In the lower right panels, $S(W_{\gamma p})$ is calculated using the leading order 
EPS09 parameterization of nuclear PDFs~\cite{eps09} extracted from the global QCD fit to available 
data;
at the leading order, EPS09 should be coupled with the CTEQ6L1 gluon distribution
of the free proton.
Note that we use EPS09 as a typical representative example---predictions for $S(W_{\gamma p})$
can also be made using other parameterizations of nuclear PDFs obtained using the global QCD 
fits~\cite{Hirai:2007sx,deFlorian:2003qf,deFlorian:2011fp} (see also figure~3 of ref.~\cite{Guzey:2013xba}).

One can see from figures~\ref{fig:S_pb208_Q2_2_4} and \ref{fig:S_pb208_Q2_3_0}
 that---with the exception of the case of the MNRT07 gluon distribution at $\mu^2=2.4$ GeV$^2$ 
at small $x$---the predictions of LTA~\cite{Frankfurt:2011cs} 
and of the global QCD analysis of nuclear PDFs~\cite{eps09} 
provide a very good description of the ALICE point at $x=10^{-3}$ and a fair description
of the point at $x=0.025$ (keeping in mind theoretical uncertainties and experimental 
errors)\footnote{One should point out that at $x=0.025$, the considered predictions of 
nuclear gluon shadowing 
converge (become essentially indistinguishable) and are all consistent with the suppression 
factor $S(W_{\gamma p})$ for Lead determined with large experimental errors. At the same time, 
the value of $S(W_{\gamma p})$ at $x=10^{-3}$ can in principle better discriminate between different
scenarios of nuclear shadowing.}.

Our results for the suppression factor $S(W_{\gamma p})$ can also be presented in the form of the 
$Pb Pb \to Pb Pb J/\psi$ cross section of exclusive photoproduction of $J/\psi$ in symmetric PbPb UPCs  
 (see the discussion in~\cite{Guzey:2013xba}). 
Combing Eqs.~(\ref{csupc}), (\ref{csa1}) and (\ref{eq:S_2}), we obtain
\begin{eqnarray}
\frac{d \sigma_{AA\to AAJ/\psi}(y)}{dy} & = & 
N_{\gamma/A}(y)\, S^2(y) \,\frac{d\sigma_{\gamma p\rightarrow J/\psi p}(y,t=0)}{dt}
\Phi_A(y) \nonumber\\
&+&N_{\gamma/A}(-y)\, S^2(-y) \frac{d\sigma_{\gamma p\rightarrow J/\psi p}(-y,t=0)}{dt}
\Phi_A(-y) \,.
\label{csupc_2}
\end{eqnarray}
In eq.~(\ref{csupc_2}), 
the first term is evaluated at 
$W_{\gamma p}(y)=2 \sqrt{\omega_H E_A}=\sqrt{2 M_{J/\psi} E_A}e^{y/2}$, 
where $E_A$ is the energy of the nuclear beam per nucleon ($E_A=1.38$ TeV for symmetric PbPb collisions 
at $\sqrt{s_{NN}}=2.76$ TeV); 
the second term is evaluated at 
$W_{\gamma p}(-y)=2 \sqrt{\omega_L E_A}=\sqrt{2 M_{J/\psi} E_A}e^{-y/2}$.

\begin{figure}[h]
\centering
\epsfig{file=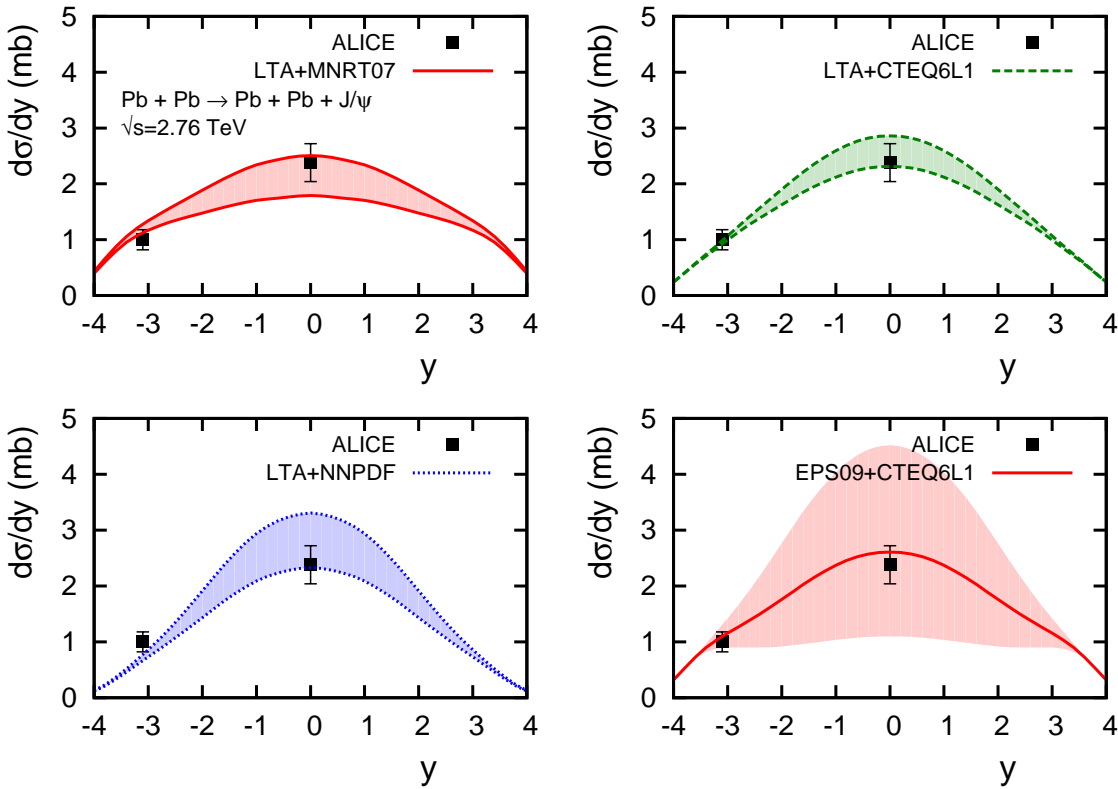,scale=1.5}
\caption{The $Pb Pb \to Pb Pb J/\psi$ cross section of exclusive $J/\psi$ photoproduction 
in symmetric PbPb UPCs as a function of the rapidity $y$ of $J/\psi$ at $\sqrt{s_{NN}} = 2.76$ TeV. 
The ALICE data points at $y=-3.1$~\cite{alice2} and
$y=0$~\cite{alice1} are compared with the LO pQCD predictions of eq.~(\ref{csupc_2})
at $\mu^2=3$ GeV$^2$.
}
\label{fig:dsdy_pb208_Q2_3}
\end{figure}

Our predictions for the $d\sigma_{AA\to AAJ/\psi}(y)/dy$ cross section of eq.~(\ref{csupc_2}) 
as a function of the rapidity $y$ of $J/\psi$ at $\sqrt{s_{NN}} = 2.76$ TeV
are presented in figure~\ref{fig:dsdy_pb208_Q2_3}.
In  eq.~(\ref{csupc_2}), the factor $S(W_{\gamma p})$ and the cross section 
$d\sigma_{\gamma p\rightarrow J/\psi p}/dt$ are evaluated at $\mu^2=3$ GeV$^2$ (see the discussion above).
One can see from the figure that the LO pQCD formalism 
provides a good description of the two ALICE data points at $y=-3.1$~\cite{alice2} and
$y=0$~\cite{alice1}.

At the LHC, ALICE, ATLAS, the CMS and LHCf experiments are equipped with zero degree calorimeters (ZDC)
allowing one to detect very forward particles produced in beam fragmentation.
This allows one 
to study coherent exclusive photoproduction of $J/\psi$ in nucleus--nucleus 
UPCs which is accompanied by electromagnetic excitation of the nuclei with the 
subsequent emission of a number of neutrons by one or by both nuclei~\cite{rsz}.
One can distinguish several channels: 1n1n corresponds to the emission of one neutron by each ion; 
XnXn -- to the emission of several neutrons; 0n1n and 0nXn -- to the excitation of only one of the 
ions; 0n0n is the case when neither of the ions emitted a neutron.
 The partial cross section for each channel is calculated 
using eq.~(\ref{csupc_2}) with the modified flux of equivalent photons corresponding to the given 
channel~\cite{rsz}. 

\begin{figure}[h]
\centering
\epsfig{file=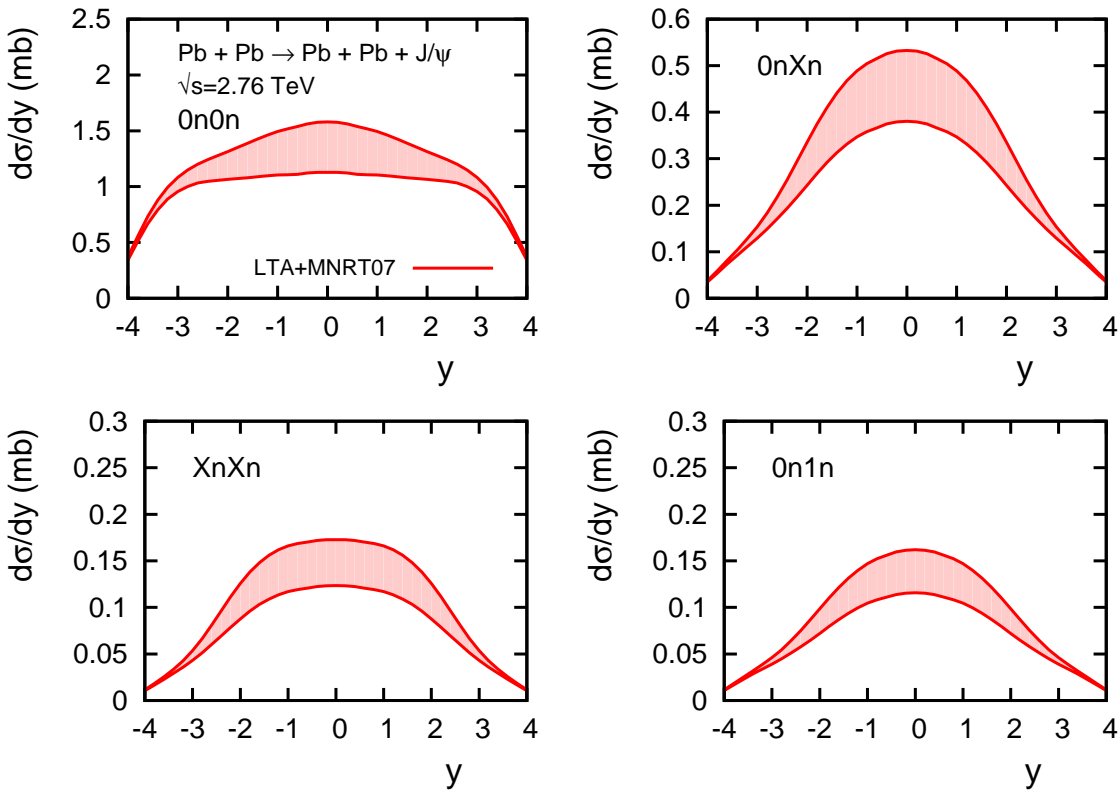,scale=1.5}
\caption{The partial $Pb Pb \to Pb Pb J/\psi$ cross sections of exclusive $J/\psi$ photoproduction 
in PbPb UPCs at $\sqrt{s_{NN}} = 2.76$ TeV accompanied by neutron emission (see text for details). 
The curves are LO pQCD predictions using the leading twist nuclear shadowing and the MNRT07 gluon 
distribution
at $\mu^2=3$ GeV$^2$.
}
\label{fig:dsdy_pb208_neutrons}
\end{figure}

Figure~\ref{fig:dsdy_pb208_neutrons} presents our predictions for the partial cross sections
of the 0n0n, 0nXn, XnXn and 0n1n channels of the $Pb Pb \to Pb Pb J/\psi$ reaction with neutron
emission at $\sqrt{s_{NN}} = 2.76$ TeV. The curves are 
LO pQCD predictions using the leading twist nuclear shadowing and the MNRT07 gluon 
distribution
at $\mu^2=3$ GeV$^2$.

In summary, using various sets of gluon distributions in the proton and predictions for the effect
of nuclear gluon shadowing,
we demonstrated 
that 
the leading logarithmic 
approximation provides the good and simultaneous description of the high-energy data on
the $\gamma p \to J/\psi p$ and  $\gamma Pb \to J/\psi Pb$ cross sections 
(see figures~\ref{fig:Sigma_proton_pQCD_Q2_3_0} and \ref{fig:S_pb208_Q2_3_0}) and on the cross section of 
exclusive $J/\psi$ photoproduction in PbPb UPCs (see figure~\ref{fig:dsdy_pb208_Q2_3}).

\section{Discussion and conclusions}

Using the leading order pQCD, we made predictions 
for the $\gamma p \to J/\psi p$, $\gamma Pb \to J/\psi Pb$ and $Pb Pb \to  Pb Pb J/\psi $
 cross sections employing various
parameterizations of the gluon distribution in the proton and Pb and compared them to the available 
LHC data on proton--proton and PbPb UPCs.
We demonstrated that the {\it simultaneous} description of the proton and Pb data is
possible and requires sizable nuclear gluon shadowing. 
We found the following constraints on 
 the description of the $\gamma p \to J/\psi p$ and  $\gamma Pb \to J/\psi Pb$
cross sections within the LO pQCD formalism: (i) the probed 
 effective scale is $\mu^2 \approx 3$ GeV$^2$ (for many sets of contemporary
gluon distributions in the proton) and (ii) the suppression factor parameterizing a host of effects
beyond the used approximation is insignificant,
$F^2(\mu^2) \approx 0.5$.
Thus, within the used formalism,
our analysis helps to place additional constraints on
the gluon distribution in the proton down to $x=6 \times 10^{-6}$ and on the gluon distribution 
in Pb down to $x=10^{-3}$. 
This is an order of magnitude improvement in the small $x$ kinematic coverage
compared to that of lepton--proton DIS at HERA and of fixed-target nuclear DIS, respectively.

Note that while we determine the value of $\mu^2$ from a comparison to the data,
one can still offer its physical interpretation. The fact that we found that $\mu^2 > M_{J/\psi}^2/4 > m_c^2$
means that in addition to  the charm quark mass $m_c$, the
transverse momentum of the charm quarks in the $J/\psi$ wave function, $k_t$, also gives a noticeable
contribution to $\mu^2$.

We also make predictions for the cross section of
coherent exclusive photoproduction of $J/\psi$ in nucleus--nucleus 
UPCs accompanied by the electromagnetic excitation of nuclei and the subsequent neutron emission.

Our findings can be compared with the results of other approaches to
exclusive photoproduction of $J/\psi$ in proton--proton and 
nucleus--nucleus UPCs at the LHC available in the literature.

(i)
Using the framework that is similar to the one used in our paper, i.e., LO pQCD, 
Adeluyi and Bertulani~\cite{ba2} (see also~\cite{Adeluyi:2011rt,Adeluyi:2012ds,Adeluyi:2013tuu})
evaluated the $\gamma p \rightarrow J/\psi p$ cross section using the MSTW08LO gluon density at 
$\mu^{2}=M^2_{J/\psi}/4=2.4$ GeV$^2$. The additional normalization factor 
of $\zeta_{V} =1/3.5=0.286$ taking into account effects beyond the used approximation
had to be introduced to explain the normalization of the 
$\gamma p \rightarrow J/\psi p$ cross section~\cite{Adeluyi:2011rt}.
However, as we explained above (see figures~\ref{fig:Sigma_proton_pQCD_Q2_2_4} and 
\ref{fig:Sigma_proton_pQCD_Q2_3_0}), the growth of the MSTW08LO gluon distribution at small $x$ 
is too steep (see figure~\ref{fig:Glue_proton_pQCD_Q2_3_comparison} of the Appendix), which means that the corresponding pQCD expression fails to simultaneously describe 
the HERA and LHCb data on the  $W_{\gamma p}$ 
behavior of the $\gamma p \rightarrow J/\psi p$ cross section. 

In conjunction with the nuclear gluon shadowing factor
$R(x,\mu^{2})$ obtained in the global QCD analyses of nuclear PDFs (HKN07, EPS08 and EPS09),
the same gluon distribution (MSTW08LO)  
was also used to calculate the  $\gamma A \rightarrow J/\psi A$ cross section~\cite{ba2}.
Such a procedure violates the consistency of the LO pQCD calculations
in the following sense. 
The 
HKN07 nuclear PDFs~\cite{Hirai:2007sx}  are based on the MRST98 parton distributions of the 
free nucleon~\cite{Martin:1998sq};
the EPS09 and EPS08 LO nPDF analysis -- on the CTEQ6L1 free nucleon PDFs. 
Hence, 
to be consistent, the use of the MSTW08 gluon density in the proton in conjunction with the HKN07, EPS08 and EPS09 gluon shadowing factors 
requires introduction of  
large $x$-dependent corrections ($x$-dependent normalization factors given by the
inversed square of  ratio of the gluon distributions presented in 
figure~\ref{fig:Glue_proton_pQCD_Q2_3_comparison} of the Appendix), 
which were not included in  
the predictions for the  $Pb Pb \to Pb Pb J/\psi$ cross section presented in ref.~\cite{ba2}.

(ii) Predictions for the $Pb Pb \to Pb Pb J/\psi$ cross section were also made 
using the framework of the color dipole model~\cite{Goncalves:2011vf,Lappi:2013am}.
The resulting calculations predict too little nuclear gluon shadowing and, as a result, overestimate
the $Pb Pb \to Pb Pb J/\psi$ cross section, see the discussion in~\cite{alice1} and 
\cite{Guzey:2013xba}. Qualitatively, these results can be readily understood: on average, 
the elementary dipole--nucleon cross section for  the $\gamma p \to J/\psi p$ process is rather small
due to squeezing of the dipole by the $J/\psi$ wave function. 
Hence, multiple rescatterings of this dipole on the target nucleons constitute essentially a higher 
twist correction to the impulse approximation and, hence, 
cannot lead to significant nuclear attenuation (shadowing).

(iii) Exclusive coherent $J/\psi$ photoproduction in heavy ion UPCs at LHC energies was also considered
in ref.~\cite{Cisek:2012yt} using the dipole formalism with the $k_T$-unintegrated gluon distribution.
In the case of the dipole--nucleus scattering, besides the Glauber-type rescattering on the target
nucleons, the approach also
takes into account the small-$x$ evolution of dipoles, i.e., gluon-fusion corrections associated with 
the rescattering of the $QQg$ dipole. 
The net resulting nuclear shadowing correction for the process and kinematics of interest 
is similar to our result 
(the shadowing correction predicted in~\cite{Cisek:2012yt} is $15-20$\% smaller than our prediction).

(iv) In the framework of the dipole formalism, an attempt to take into account the contribution of a 
broad range of dipole sizes, i.e., the contribution of both perturbative and non-perturbative 
QCD contributions,
to the $\gamma p \to J/\psi p$ cross section was made 
in~\cite{Frankfurt:1995jw,Frankfurt:1997fj,Frankfurt:2000ez}.
It was estimated that the 
$\gamma p \to J/\psi p$ process probes the effective scales $\mu^2 > 2.4$ GeV$^2$ and that the relativistic 
corrections are large, i.e., the factor of $F^2$ is small. 
While the latter 
differs from
our finding that
$F^2 \approx 0.5$ which is consistent with the analysis of~\cite{Ryskin:1995hz},
one has to keep in mind that the gluon distributions used in different analyses are different which
naturally results in different $F^2$ factors. 
 The extension of the analysis
to the nuclear case predicts sizable suppression of the $\gamma A \to J/\psi A$ cross section due
to the gluon nuclear shadowing~\cite{Frankfurt:1995jw}.

In the near future, there will be available results of measurements of $J/\psi$ photoproduction
from the proton--lead run at $\sqrt{s_{NN}}=5.02$ TeV.
The new data will mostly probe the photon--proton interaction, which, in turn, will influence
the precision of the analysis of the photon--nucleus cross 
section~\cite{Guzey:2013taa}.

\section*{Acknowledgements}

The authors would like to thank M.~Ryskin and M.~Strikman for illuminating discussions of 
their respective results on photoproduction of $J/\psi$ on the proton.

\section*{Appendix: Leading order gluon distributions in the proton and corresponding $\alpha_s$ used
in the present work}
\label{sec:Appendix_A}

\begin{figure}[h]
\centering
\epsfig{file=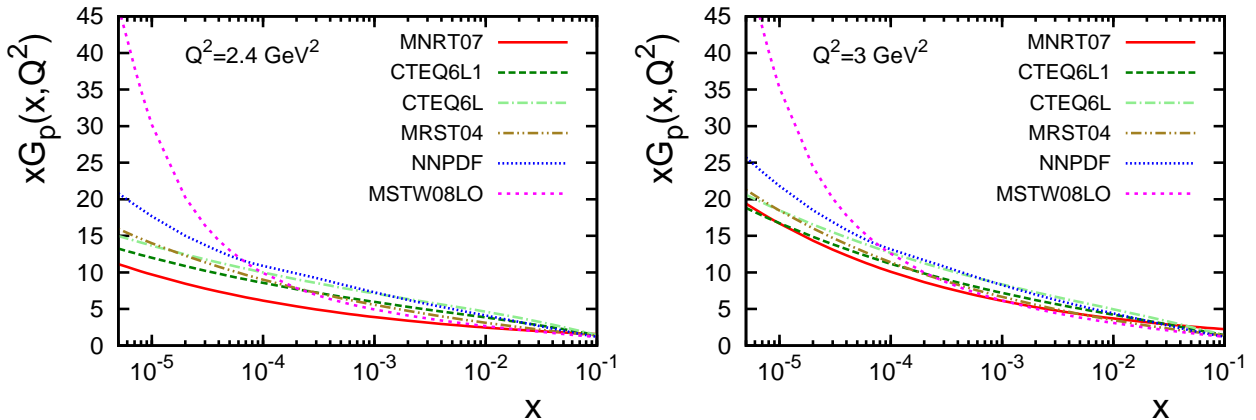,scale=1.3}
\caption{Leading order gluon parton distributions in the nucleon at $Q^2=2.4$ GeV$^2$ (left) and
at $Q^2=3$ GeV$^2$ (right) used in this work.}
\label{fig:Glue_proton_pQCD_Q2_3}
\end{figure}

For the calculation of the perturbative $\gamma p \to J/\psi p$ cross section, 
see Eqs.~(\ref{dcslo}) and (\ref{csprot}), we used six different leading order (LO)
gluon distributions in the proton: MNRT07~\cite{Martin:2007sb}, 
CTEQ6L1~\cite{Pumplin:2002vw}, CTEQ6L~\cite{Pumplin:2002vw}, MRST04~\cite{Martin:2006qz}, NNPDF~\cite{Ball:2010de} 
and MSTW08LO~\cite{Martin:2009iq}. (Note that these PDFs are listed in the order 
of an increasing magnitude of $xG_p(x,Q^2)$ at $x=5 \times 10^{-6}$ and $Q^2=2.4$ GeV$^2$.) 
Figure~\ref{fig:Glue_proton_pQCD_Q2_3} shows these parameterizations for 
$xG_p(x,Q^2)$ as a function of $x$ at $Q^2=2.4$ GeV$^2$ (left panel) and at $Q^2=3$ GeV$^2$ 
(right panel).

Table~\ref{table:alpha_s} summarizes the values of strong coupling constant $\alpha_s(\mu^2)$ 
corresponding to the examined gluon parton distributions, which we used in our numerical 
predictions using eqs.~(\ref{csprot}), (\ref{csa1}), and (\ref{csupc_2}).

\begin{table}[htbp]
  \centering
  \begin{tabular}{|c|c|c|c|c|}
    \hline
Parameterization & $\alpha_s(Q^2=2.4 \ {\rm GeV}^2)$ & $\alpha_s(Q^2=3 \ {\rm GeV}^2)$ &
$\alpha_s(M_Z^2)$ & Comments \\
\hline
MNRT07~\cite{Martin:2007sb}   & $0.295$ & $0.282$ & $0.118$  & $\Lambda_4=120$ MeV \\
CTEQ6L1~\cite{Pumplin:2002vw} & $0.382$ & $0.361$ & $0.130$  & $\Lambda_4=215$ MeV \\
CTEQ6L~\cite{Pumplin:2002vw}  & $0.330$ & $0.314$ & $0.118$  & NLO with $\Lambda_4=326$ MeV \\
MRST04~\cite{Martin:2006qz}   & $0.386$ & $0.365$ & $-$      & $\Lambda_4=220$ MeV \\
NNPDF~\cite{Ball:2010de}      & $0.301$ & $0.289$ & $0.119$  & $\Lambda_4=127$ MeV \\
MSTW08LO~\cite{Martin:2009iq} & $0.480$ & $0.448$ & $0.139$  & $\Lambda_4=322$ MeV \\
\hline
\end{tabular}
\caption{The running strong coupling constant $\alpha_s(\mu^2)$ corresponding to the examined 
gluon parton distributions.}
\label{table:alpha_s}
\end{table}

\begin{figure}[h]
\centering
\epsfig{file=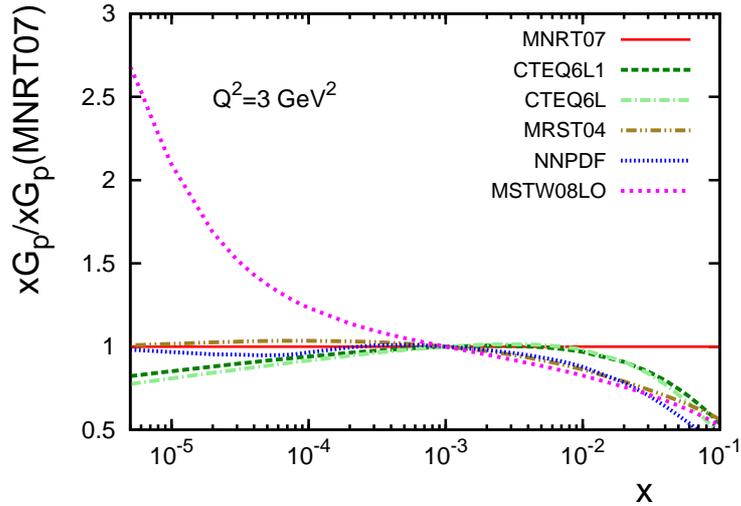,scale=1.3}
\caption{The ratios of the studied LO gluon 
distributions of the proton to the MNRT07 gluon distribution as a function of $x$ at
$Q^2=3$ GeV$^2$. The ratios are normalized to be equal to unity at $x=10^{-3}$.}
\label{fig:Glue_proton_pQCD_Q2_3_comparison}
\end{figure}

Figure~\ref{fig:Glue_proton_pQCD_Q2_3_comparison} shows the ratios of the studied gluon 
distributions of the proton to the MNRT07 gluon distribution as a function of $x$ at
$Q^2=3$ GeV$^2$. The ratios are normalized to be equal to unity at $x=10^{-3}$. 

We explained in section~\ref{subsec:pQCD_proton} that to evaluate the factors of $\eta$ (the ratio of the real
to the imaginary parts of the $\gamma p \to J/\psi p$ amplitude) and $R_g$ (the skewness factor), 
we used the fit $xG_p (x,\mu^2) \propto 1/x^{\lambda}$ in the interval $5 \times 10^{-6} \leq x \leq 0.01$.
The resulting values of $\lambda$ are summarized in table~\ref{table:lambda}.

Note that the MSTW08LO gluon distribution cannot be fitted using a single $\lambda$ in 
the entire $5 \times 10^{-6} \leq x \leq 0.01$ interval. To achieve an acceptable accuracy
of the  $xG_p (x,\mu^2) \propto 1/x^{\lambda}$ fit, we allowed $\lambda$ to vary with $x$: the last line 
of table~\ref{table:lambda} gives the range of obtained values; $\lambda$ decreases with an 
increase of $x$.

\begin{table}[htbp]
  \centering
  \begin{tabular}{|c|c|c|}
    \hline
Parameterization & $\lambda(Q^2=2.4 \ {\rm GeV}^2)$ & $\lambda(Q^2=3 \ {\rm GeV}^2)$ \\
\hline
MNRT07~\cite{Martin:2007sb}   & $0.20$ & $0.22$ \\
CTEQ6L1~\cite{Pumplin:2002vw} & $0.16$ & $0.19$ \\
CTEQ6L~\cite{Pumplin:2002vw}  & $0.15$ & $0.18$ \\
MRST04~\cite{Martin:2006qz}   & $0.20$ & $0.22$ \\
NNPDF~\cite{Ball:2010de}      & $0.20$ & $0.22$ \\
MSTW08LO~\cite{Martin:2009iq} & $0.62-0.30$ & $0.58-0.30$ \\
\hline
\end{tabular}
\caption{The parameter $\lambda$ in the fit $xG_p (x,\mu^2) \propto 1/x^{\lambda}$ of the studied proton
gluon distributions in the interval $5 \times 10^{-6} \leq x \leq 0.01$ at $\mu^2=2.4$ GeV$^2$ and 
 $\mu^2=3$ GeV$^2$.}
\label{table:lambda}
\end{table}

\end{document}